\documentclass[conference]{IEEEtran}
\IEEEoverridecommandlockouts
\usepackage{cite}
\usepackage{amsmath,amssymb,amsfonts}
\usepackage{algorithmic}
\usepackage{graphicx}
\usepackage{textcomp}
\usepackage{xcolor}
\def\BibTeX{{\rm B\kern-.05em{\sc i\kern-.025em b}\kern-.08em
    T\kern-.1667em\lower.7ex\hbox{E}\kern-.125emX}}

\usepackage{color}
\usepackage{minted}
\usepackage{listings}
\usepackage[T1]{fontenc}
\usepackage{makecell}
\usepackage{booktabs}
\usepackage{tabularx} 

\usepackage{soul} 
\soulregister{\cite}{1}
\soulregister{\ref}{1}
\soulregister{\textbf}{1}
\soulregister{\textit}{1}
\usepackage{enumitem} 
\usepackage{multirow}

\colorlet{punct}{red!60!black}
\definecolor{background}{HTML}{FFFFFF}
\definecolor{delim}{RGB}{20,105,176}
\colorlet{numb}{magenta!60!black}
\definecolor{dkgreen}{rgb}{0,0.6,0}
\newcommand{\nehacomment}[1]{}
\usepackage{colortbl}
\usepackage{enumitem}

\renewcommand{\lstlistingname}{API}

\lstdefinelanguage{json}{
    basicstyle=\fontsize{7}{8}\selectfont\ttfamily,
    showstringspaces=false,
    breaklines=true,
    frame=single,
    backgroundcolor=\color{background},
    literate=
     *{0}{{{\color{numb}0}}}{1}
      {1}{{{\color{numb}1}}}{1}
      {2}{{{\color{numb}2}}}{1}
      {3}{{{\color{numb}3}}}{1}
      {4}{{{\color{numb}4}}}{1}
      {5}{{{\color{numb}5}}}{1}
      {6}{{{\color{numb}6}}}{1}
      {7}{{{\color{numb}7}}}{1}
      {8}{{{\color{numb}8}}}{1}
      {9}{{{\color{numb}9}}}{1}
      {:}{{{\color{punct}{:}}}}{1}
      {,}{{{\color{punct}{,}}}}{1}
      {\{}{{{\color{delim}{\{}}}}{1}
      {\}}{{{\color{delim}{\}}}}}{1}
      {[}{{{\color{delim}{[}}}}{1}
      {]}{{{\color{delim}{]}}}}{1},
}
\begin{document}

\title{{EdgeFaaS: A Function-based Framework for Edge Computing}\\
\thanks{This work is partly supported by NSF award No. 2231874 and National Institutes of Health AIM-AHEAD Program Agreement No. 1OT2OD032581.}
}

\author{\IEEEauthorblockN{Neha Vadnere, Yu-Ting Wang, Yitao Chen, Sreehari Sadesh, Ming Zhao}
\IEEEauthorblockA{\textit{Arizona State University}, Tempe, USA \\
}
}

\maketitle

\thispagestyle{plain}
\pagestyle{plain}

\begin{abstract}
Edge computing brings unique challenges as the resources on the edge are highly diverse in capabilities and capacities, and highly distributed across many users and the physical world. Existing distributed computing frameworks cannot adequately handle this level of heterogeneity and distribution.
This paper proposes EdgeFaaS, a novel function-based edge computing framework to enable edge applications to effectively utilize heterogeneous resources distributed across the Internet of Things (IoT), edge, and cloud for computing. It proposes function virtualization and storage virtualization to abstract distributed and heterogeneous physical resources and provides consistent virtual interfaces for deploying and executing functions and storing and accessing data. EdgeFaaS provides comprehensive support to diverse edge computing workflows, and at the same time allows users to flexibly adjust the configurations and explore various important tradeoffs. To demonstrate its usability, the paper also presents the implementation and evaluation of three representative workflows on EdgeFaaS for video analytics, federated learning, and audio classification, on a real testbed of 100+ geographically distributed IoT devices, edge servers, and cloud services. 
EdgeFaaS allows users to flexibly explore the deployment configurations of these workflows over distributed and heterogeneous resources. For example, users can easily vary the function placement of the video processing pipeline across IoT, edge, and cloud resources and study the tradeoff between computation and communication costs; users can also flexibly adjust the cluster count and size in the hierarchical federated learning system and explore the tradeoff between training accuracy and speed. 

\end{abstract}

\begin{IEEEkeywords}
FaaS , Edge Computing, Service, Workflow
\end{IEEEkeywords}

\section{Introduction}
\label{sec:introduction}

While the Internet-of-Things (IoT) revolution brings unprecedented opportunities for economic growth, it also presents important challenges to our existing computational infrastructure. The cloud is projected to fall short by two orders of magnitude to either transfer, store, or process such vast amounts of streaming data. Moreover, the cloud-based solution will not be able to provide a satisfactory quality of experience for many time-sensitive or privacy-sensitive IoT applications. A vast amount of computation and storage resources are being deployed in proximity to IoT devices, enabling a new computing paradigm, namely ``edge computing''. 

Compared to traditional computing paradigms, edge computing has distinct characteristics. Resources on the edge are \textit{highly diverse} in capabilities and capacities and \textit{highly distributed} across IoTs, edge servers, and cloud data centers. 
Edge applications are driven by streams of \textit{dynamic} data sensed from users and the physical world and need to act upon the data and respond to events that occur to the users and in the physical world in a \textit{timely} fashion. 
Existing distributed computing frameworks are insufficient to handle this level of heterogeneity, distribution, and dynamism.

This paper proposes \textbf{EdgeFaaS}, a novel function-based edge computing framework to enable dynamic data-driven edge applications to effectively utilize heterogeneous resources distributed across the IoT, edge, and cloud for function executions and data storage.

First, EdgeFaaS allows cluster resources and individual servers and devices to be managed under the same framework and enables users to utilize these resources using the powerful abstraction of Function as a Service (FaaS). 
FaaS supports fine-grained application development, deployment, and scaling, and is the latest service model of cloud computing~\cite{openfaas,AWSLambda1}. Recently, FaaS has also extended to edge computing~\cite{wolski2019cspot,nanolambda,zhang2020tinyedge}, but existing solutions have limitations in supporting the high diversity in resources across IoT, edge, and cloud.
EdgeFaaS addresses these limitations by consolidating these highly diverse and distributed resources as a single, unified FaaS resource pool and enabling users to write their functions once and deploy and run them anywhere (with sufficient resources).

Second, EdgeFaaS provides novel function and storage virtualization, offering a unified interface for users to conveniently and productively utilize an edge system's heterogeneous and distributed computing and storage resources. With such virtualization, the locations of functions and data and the interfaces of the underlying FaaS and storage resources are entirely hidden from users; users deploy and invoke/access their functions/data through EdgeFaaS. This frees users from the burden of interacting with heterogeneous and distributed resources, which may have different interfaces, deployment models, and hardware capabilities, and allows them to access these resources through EdgeFaaS consistently.

Third, EdgeFaaS allows diverse edge applications to be flexibly specified and deployed as workflows of functions, by extending the Serverless Workflow Specification standard~\cite{cncf_serverless_workflows}) to support edge workflows.
It automatically optimizes data and function placement by taking into account factors such as data locality and function dependency to minimize data access/transfer latency in edge applications, and data privacy to ensure that data is only stored and accessed on user-trusted resources. 
It also allows users to flexibly explore workflow configurations in order to optimize application-specific objectives such as speed and accuracy.

We implemented and evaluated three representative edge workflows to demonstrate the capability and usability of EdgeFaaS.
The video analytics workflow 
represents applications with multiple stages for processing IoT data in a pipeline fashion. 
The hierarchical federated learning workflow 
represents applications that need to process the data from many IoT devices and then aggregate their results. It also models applications that have data privacy requirements and allow data to be stored and processed only on certain trusted resources. 
The audio classification workflow models typical machine learning based applications that train a generalized model in the cloud, fine-tune specific versions of the model on the edge, and deploy the fine-tuned models on IoTs for inference.
EdgeFaaS allows users to flexibly explore the configurations of these workflows over distributed and heterogeneous resources, such as the placement of various stages in the video analytics pipeline, and optimize the various important tradeoffs such as the model accuracy vs. training speed of federated learning. 

In summary, the contributions of EdgeFaaS are as follows: 1) It provides a function-based computing framework that unifies IoTs, edge, and cloud resources and allows users to conveniently deploy and execute their functions across the resources; 2) It employs novel function virtualization and storage virtualization to address the resource heterogeneity in software and hardware; and 3) It provides comprehensive support for diverse edge computing workflows and enables users to conveniently deploy the workflows onto distributed, heterogeneous resources and flexibly explore and optimize workflow configurations.

\section{Background and Related Works}
\label{sec:background}

A proper abstraction is key to enabling computing and data access across the heterogeneous compute and storage resources on the edge and across the multiple tiers of resources from edge to cloud. We advocate the use of functions to provide this abstraction. Function-as-a-Service (FaaS) is a service-oriented computing model that allows users to run functions without the complexity of building and maintaining the infrastructure. 
Users define functions, and FaaS deploys applications in the unit of sandboxed functions. Functions can be invoked by events or other functions; different functions can communicate and coordinate to form a workflow. FaaS quickly and dynamically scales a function by spawning the function sandboxes on demand. As soon as a request finishes, its function sandboxes can be shut down to release resources, and no state is guaranteed to be preserved.

Compared to other service models such as IaaS (Infrastructure-as-a-Service) and conventional PaaS (Platform-as-a-Service), FaaS has several important benefits. 
First, FaaS simplifies application development by allowing users to focus on the core functionalities of their applications and delegate function deployment and invocations to the FaaS framework.
Second, FaaS enables faster initialization by deploying and launching functions with smaller image sizes than entire applications. 
Third, FaaS improves resource utilization by providing resources to individual functions on demand, instead of entire applications.

AWS IoT Greengrass~\cite{AWSGreengrass} and Azure IoT Edge~\cite{AzureIoTEdge} enable users to run functions on edge devices.
However, as mainly an extension of their corresponding cloud services to the edge, such solutions have several limitations.
First, they require the use of a specific software stack and their use is often tied to a specific cloud backend, which limits their support of heterogeneity. For example, to use AWS IoT Greengrass, users need to run its various components on IoTs and edge devices, under the management of AWS cloud IoT Core. 
Second, unlike the cloud which provides users great location transparency, AWS IoT Greengrass requires users to specify which devices to deploy their functions and how to connect them. This essentially ties the functions to specific devices and prevent the users from enjoying the benefits of elasticity.

Recent position papers~\cite{runyu_edge23,Edgeless,serverless19icfc} have identified the opportunities and challenges as well as architecture of function-based edge computing.
Baresi et al. also implemented a primitive, OpenWhisk-based serverless prototype for edge computing, but it did not address the heterogeneity of resources and the need of consistent data access across IoT, edge, and cloud resources~\cite{serverless19icfc}.
Related works have addressed various aspects of this novel computing paradigm. 
CSPOT~\cite{wolski2019cspot} utilizes Linux containers for function executions and provides an append-only object store; NanoLambda~\cite{nanolambda} further integrates a lightweight Python runtime for function executions on
weak IoT devices such as microcontrollers. 
CWASI-shim~\cite{cwasi} targets resource-constrained IoTs by running functions inside lightweight WebAssembly sandboxes.

EdgeFaaS complements these related works by supporting diverse function execution environments and frameworks and the composition and execution of versatile function workflows across edge and cloud resources. Moreover, EdgeFaaS provides cross-tier data stores with commonly used interface, enabling consistent views and convenient use of data across the edge system.

While function scheduling is not the focus of this paper, 
EdgeFaaS can serve as the foundational infrastructure to integrate advanced scheduling algorithms.
FaDO~\cite{smith2022fado} presented several load-balancing algorithms for serverless edge-cloud clusters; 
DFaaS~\cite{9657141}, 
E Carlini \textit{et al}~\cite{carlini2025dynamic} and DMSA~\cite{chen2026dmsa} proposed a decentralized architecture for balancing the traffic load across edge nodes~\cite{waterfall};
and NanoLambda~\cite{nanolambda} also proposed a scheduler for latency-aware function scheduling.

\section{Architecture and Design}
\label{sec:arch}

\begin{figure}[h]
	\centering
	\includegraphics[width=0.95\columnwidth]{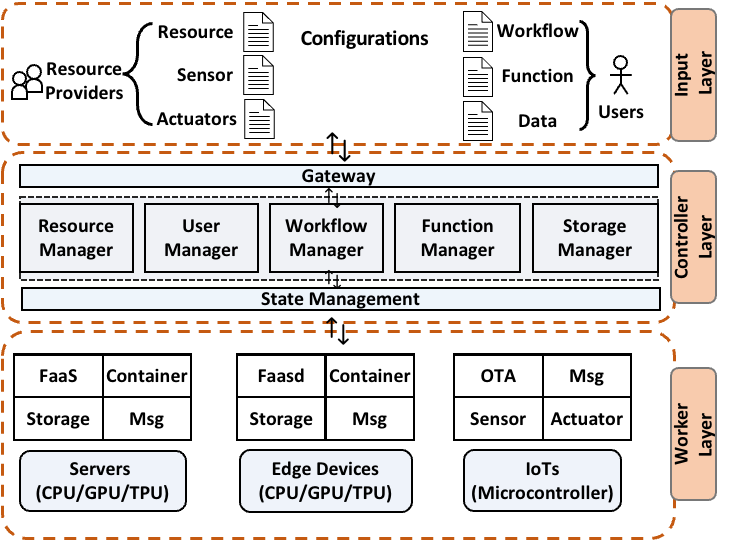}
    \vspace{-6pt}
	\caption{\small EdgeFaaS system architecture.
    }
    \vspace{-12pt}
	\label{fig:EdgeFaaS-system-design}
	\vspace{-1pt}
	\centering
\end{figure}

EdgeFaaS follows a three-layer architecture comprising the \textit{Input Layer}, \textit{Controller Layer}, and \textit{Worker Layer}, as illustrated in Fig.~\ref{fig:EdgeFaaS-system-design}.
The \textit{Input Layer} allows users and resource providers to configure resources and applications. Resource providers specify resource details via configuration files, while users provide function code (as zip files or container images) and workflow specifications.
The \textit{Gateway} provides a virtual interface with unified REST APIs to resource providers for resource management and to users for deploying/executing functions and storing/accessing data.
The \textit{Controller Layer} employs distributed managers to manage the heterogeneous resources and the workflow/function execution and data access/storage on the resources.  
The \textit{Worker Layer} provides the environment to execute functions and store data over heterogeneous, geographically distributed resources, including cloud and edge servers, edge devices, and IoTs.

\vspace{3pt}
\subsection{Resource Virtualization}
\label{sec:resource}

EdgeFaaS manages heterogeneous resources across different tiers, providing computing and storage capabilities for deploying and running functions and storing and accessing data.
There are mainly three broad categories of function resources. The first category is \textit{FaaS clusters and servers} such as cloud FaaS services (e.g., Amazon Lambda~\cite{AWSLambda1}, Google Cloud Functions~\cite{googlefunctions}, Azure Functions~\cite{azurefunction}) and edge servers with standalone FaaS systems (e.g., OpenFaaS~\cite{openfaas}, OpenWhisk~\cite{OpenWhisk1}, faasd~\cite{faasd}) which provide high-performance function resources with CPUs, GPUs, and TPUs. 
A FaaS cluster or server typically exposes its resources for function deployment and execution via a gateway and specific APIs.

The second category is \textit{edge devices} such as Raspberry Pi, NVIDIA Jetson, Coral Edge TPU which provide edge-optimized but limited CPUs and accelerators for executing only one or a few functions. 
Such devices run lightweight FaaS platforms (e.g., faasd~\cite{faasd}, tinyFaaS~\cite{tinyfaas2020}), and expose their resources to EdgeFaaS through their gateways and APIs.

The third category is \textit{IoTs} equipped with only microcontrollers such as Raspberry Pico and ESP32 which can run only code that is flashed onto the device firmware. Microcontrollers cannot run any FaaS frameworks, nor have the ability to dynamically switch between the execution of different functions. Instead, EdgeFaaS leverages the microcontrollers' over-the-air (OTA) update capability to download function code over the network and automatically flash the firmware. 

In addition, storage resources are typically provided by cloud storage services (e.g., Amazon S3~\cite{s3}, Google Cloud Storage~\cite{googlestorage}, Azure Storage\cite{azurestorage}) and edge servers with standalone storage systems (e.g., MinIO~\cite{minio}).

EdgeFaaS also supports sensors that observe various phenomena, such as temperature, audio, and video, along with their metadata, including name, description, observation encoding type, and observation type. 
Data generated by these sensors is organized into collective datastreams.
EdgeFaaS presents sensors using the Open Geospatial Consortium (OGC) SensorThings standard APIs \cite{OGCSensorThings}. 

EdgeFaaS enables resource providers to register a new resource by uploading a resource description YAML file in EdgeFaaS. The description includes fields (listed in Table~\ref{table:registration-fields}) for the name, device hardware (such as CPU, GPU, TPU, microcontroller), tier type, and service endpoints for FaaS, object store, message broker, and monitoring. 
EdgeFaaS manages and virtualizes these heterogeneous, distributed compute and storage resources. As discussed in the following sections, EdgeFaaS' function virtualization and storage virtualization create the perception of a single, cohesive pool of resources and provide unified interfaces for users to deploy/execute functions and store/access data.
Internally, EdgeFaaS uses each resource's corresponding endpoints to deploy/invoke functions, object store endpoints to create/upload/download data and objects, monitoring endpoints to get resource/function telemetry, and sensor endpoints to create/list sensors and create/publish/subscribe to datastreams, and observations endpoints to create/publish/subscribe to observed phenomenon.

\begin{table}[h]
\centering
\vspace{-5pt}
\caption{Resource Description with Sample Elements}
\vspace{-5pt} 
\label{table:registration-fields}
\footnotesize 
\renewcommand{\arraystretch}{0.9} 
\setlength{\tabcolsep}{4pt} 
\begin{tabular}{lll}
\toprule
\textbf{Category} & \textbf{Field} & \textbf{Sample Element} \\
\midrule
\multirow{7}{*}{\textbf{Resource}}  & Name          & CLOUD-1 \\
                                    & Device        & CPU, GPU \\
                                    & Type          & Cloud \\
                                    & FaaS          & 107.60.105.79:8080 \\
                                    & Monitoring    & 107.60.105.79:30090 \\
                                    & Object\_store & 107.60.105.79:9000 \\
                                    & Msg broker    & 107.60.105.79:1833 \\
\addlinespace[0.8em] 
\multirow{3}{*}{\textbf{Sensor}}    & Name          & Camera \\
                                    & Description   & Raspberry Pi Camera Module 3 \\
                                    & Endpoint      & 107.60.105.79:8081 \\
\bottomrule
\end{tabular}
\end{table}
\vspace{-5pt}

\subsection{Function Virtualization}
\label{sec:function}

Function virtualization creates virtual representations of user-defined functions, abstracting their deployment location, method, and configuration from the underlying physical FaaS resources. This stems from the need of unifying the interfaces provided by different FaaS platforms and heterogeneous resources. Traditional FaaS platforms lack the ability to abstract this heterogeneity, leading to fragmented deployments and increased complexity for developers.

Function virtualization provides a consistent user interface by assigning a unique virtual URL to each function, allowing users to deploy or invoke functions using REST APIs in the consistent format \texttt{\small <workflowName>/<functionName>}. This enables access to function services without the need to know their physical locations. 
Internally, EdgeFaaS maintains the mappings from the virtual functions to their physical deployments.
Due to function virtualization, EdgeFaaS implements function deployment in two steps. \textit{Virtual function deployment} occurs when the user sends a ``Deploy'' request with the function description to EdgeFaaS for each function in the workflow. It only registers the function with EdgeFaaS without actually deploying it on any physical resource. 

The function description (in YAML/JSON format) includes details such as the function name, function-container location URL, start process, and handler. Optional parameters such as resource limits, annotations, and environment variables can also be configured. The current implementation of EdgeFaaS supports functions implemented in Python which is the \textit{de facto}  programming language for FaaS. A simple function example is shown below:

\renewcommand{\lstlistingname}{List} 
\begin{lstlisting}[language=json]
{
 "service": "face-recognition",
 "image": "dockerhubuser/face-recognition:latest",
 "envProcess": "python3 index.py
}
\end{lstlisting}

\textit{Physical function deployment} takes place when the user sends an ``Invoke'' request for the function or workflow. It enables EdgeFaaS to deploy the function on the FaaS endpoint of a physical resource. During physical deployment, EdgeFaaS evaluates the function's resource requirements, such as CPU, GPU, and memory, and selects the appropriate resource for function deployment. In addition, EdgeFaaS ensures that the data that a function depends on is accessible from the resource that it is deployed to, which is important to protect the data's privacy requirement.

After the physical resource is chosen for function deployment, EdgeFaaS uses the interface provided by the physical resource to deploy the function. For cloud FaaS services and standalone FaaS systems, EdgeFaaS contacts the gateway of the FaaS service/system and invokes the interface to deploy the function encapsulated as a zip file or container image.

For microcontrollers, EdgeFaaS utilizes over-the-air (OTA) update to deploy functions over the network.
The EdgeFaaS component running on a microcontroller provides REST APIs for downloading a function, implemented as part of a firmware update, and for invoking the function. To update the firmware, EdgeFaaS downloads the update, stores it in an OTA partition, and reboots the microcontroller from the partition. Once the microcontroller is rebooted, the newly deployed function is ready to be invoked. It employs MicroPython and a lightweight HTTP server to support Python functions and remote invocations. It utilizes a watchdog timer to reset the microcontroller if the previously deployed function becomes unresponsive or behaves unexpectedly.

\subsection{Storage Virtualization}
\label{sec:storage-management}

Storage virtualization creates a virtual representation of an object store and provides a consistent virtual interface using virtual URLs to buckets and objects, which abstract away the storage's physical location and provide a unified interface for seamless access to data. 
The virtual URL of EdgeFaaS data bucket uses \texttt{\small<workflowName>/<bucketName>} format.
Internally, EdgeFaaS maps the virtual buckets to their physical locations, and maintains a consistent view and access of the buckets wherever they are physically stored or even migrated across the distributed and heterogeneous storage resources.

EdgeFaaS follows the Data Locality principle to decide data placement, which ensures that data is always stored at or near where the function that generates the data runs in order to reduce the data transfer cost.
EdgeFaaS supports data privacy by allowing users to specify the trusted resources to store their data when creating the bucket. 

Below shows an example:

\renewcommand{\lstlistingname}{List} 
\begin{lstlisting}[language=json]
{
  "Name":"data1",
  "Namespace":"fl",
  "TrustedEndpoint":"107.60.105.79:9000"
}
\end{lstlisting}

\subsection{User Virtualization}
\label{sec:user} 

Considering heterogeneous resources managed by different FaaS and storage systems, EdgeFaaS provides users with virtual identities for consistently accessing resources across these systems: the same user can use the same virtual identity to access different resources contributed by different providers. 

Users need authentication and authorization with EdgeFaaS to deploy workflows and functions or access the object store and functions in EdgeFaaS. User authentication uses a username/password or popular OpenID Connect providers~\cite{OpenIDConnect} like Google Identity Platform, vIDM, or Auth0.
To provide users access to the physical FaaS and storage resources, EdgeFaaS maintains the mapping between users' virtual identities in EdgeFaaS and their physical identities on the resources that they have access to. EdgeFaaS uses the interfaces provided by the underlying FaaS and storage services to manage the physical users. Concurrent users on EdgeFaaS are isolated from each other as their corresponding physical users are isolated from each other on the shared resources.

\subsection{Workflow Management}
\label{sec:workflow}

EdgeFaaS provides versatile and standardized support for workflow of functions. Workflow is a sequence of functions that implements application logic by interacting and exchanging data with one another. Workflows streamline the end-to-end application logic by allowing users to specify the ordering of the functions using a descriptive configuration. 

\subsubsection{Workflow Specification}

To support diverse workflows, EdgeFaaS allows users to provide a comprehensive configuration that includes essential attributes specific to the given workflow. 
EdgeFaaS extends the Serverless Workflow Specification~\cite{cncf_serverless_workflows} set by the Cloud Native Computing Foundation (CNCF) to define workflows and support additional features that EdgeFaaS provides (an example is shown in List~\ref{listing:va-workflow-creation}). This file, represented in YAML format, provides key inputs for EdgeFaaS to understand the specifics of the workflow, such as its name, the starting function that sets the execution in motion, and the various input and output specifications required for proper functioning. Additionally, the configuration also outlines the dependencies between functions and the sequence in which they should be executed. 

\renewcommand{\lstlistingname}{List}

\begin{lstlisting}[language=json,firstnumber=1,captionpos=b, caption=Workflow configuration for a video analytics pipeline, label=listing:va-workflow-creation]
document:
  dsl: 1.0.0-alpha1
  namespace: videoanalytics
  name: videoanalytics
  version: 0.1.0
do:  
  - video-splitting:
      do:
        - handle:
            execute: true
      dependencies: null      
  - motion-detection:
      do:
        - handle:
            execute: true
      dependencies: video-splitting
  - face-extraction:
      do:
        - handle:
            execute: true
      dependencies: motion-detection
  - face-recognition:
      do:
        - handle:
            execute: true
      dependencies: face-extraction
      output: va/output
\end{lstlisting}

A user submits a new workflow request to EdgeFaaS through a REST API, providing the workflow configuration. Upon successful workflow creation, EdgeFaaS registers the workflow and generates virtual URLs for each function in the format of \texttt{\small <workflowName>/<functionName>}. 
The \texttt{\small<workflowName>} represents a workflow namespace to provide isolation to its functions and data.
EdgeFaaS parses the workflow specification into a Directed Acyclic Graph (DAG) representing functions as nodes and control-flow dependencies as edges, which serves as the blueprint for workflow execution.

\subsubsection{Workflow Orchestration}

EdgeFaaS' distributed orchestrators manage the deployment, execution, and failure detection and handling for a workflow and its functions. Orchestration operates at two levels. The \textit{Global Orchestrator} maintains a holistic view of all workflows, manages the function/data placement for each workflow, and partitions the workflow across edge orchestrators for decentralized runtime management.
\textit{Edge Orchestrators} each manages its assigned workflow partitions by managing the corresponding function executions within its own region. 
This two-level design promotes scalability as edge orchestrators independently manage the function executions of their own partitions with local state management, while global orchestrator manages only partition-level executions. 

Specifically, the global orchestrator performs function placement for a workflow by instantiating its DAG into a Deployment DAG which includes its deployment-specific information such as the endpoint bindings.
It performs function placement by considering resources that meet the functions' resource and data requirements.
It then partitions the Deployment DAG by grouping functions co-located in the same network region based on placement decisions, and dispatches each partition to its assigned Edge Orchestrator.
Edge orchestrators invoke and monitor function executions according to their assigned workflow Deployment DAG partitions.
Each edge orchestrator tracks function dependencies and queue dependent functions till all dependencies within its assigned partition are satisfied. 
Fan-in functions wait for all dependencies to complete before execution; branching functions evaluate conditions to determine downstream execution paths. 
Global orchestrator manages function dependencies between DAG partitions. For example, for a fan-in function that aggregates outputs from multiple DAG partitions, the global orchestrator tracks the status of all the partitions and triggers the fan-in execution only after receiving completion notifications from all the involved edge orchestrators.

\subsubsection{Failure Detection and Handling}

EdgeFaaS' monitoring service tracks the health status of all the underlying resources and automatically excludes the unhealthy ones from physical function deployment. However, resources can still fail after functions are deployed.
Edge orchestrators detect function execution failures based on timeouts and automatically handle them with retries. 
Functions report execution status to orchestrator-provided callback URLs upon completion. 
If an edge orchestrator does not receive callbacks within the function timeout, it attempts retries, as timeouts indicate prolonged execution, hanging, or crashes. 
After exhausting retries, the edge orchestrator marks the function execution as failed, and reports the partition failure to the global orchestrator. 
EdgeFaaS automatically redeploys functions affected by failed resources and the user can reinvoke the workflow execution.

\section{Use Case Workflows}
\label{sec:workflow_examples}

In this section, we present three specific edge workflows implemented using EdgeFaaS to showcase its capability and usability: an end-to-end video analytics workflow, a federated learning workflow, and an audio classification workflow, all of which are representative of edge applications and yet have distinct characteristics in terms of workflow composition and resource demands.
EdgeFaaS not only allows users to conveniently deploy and execute these diverse workflows on distributed, heterogeneous resources, but also enables users to flexibly adjust key configurations and explore important tradeoffs that are specific to each workflow.

\subsection{Video Analytics Workflow} 
\label{sec:video-analytics-workflow}

\begin{figure}[h]
	\centering
	\includegraphics[width=0.75\columnwidth]{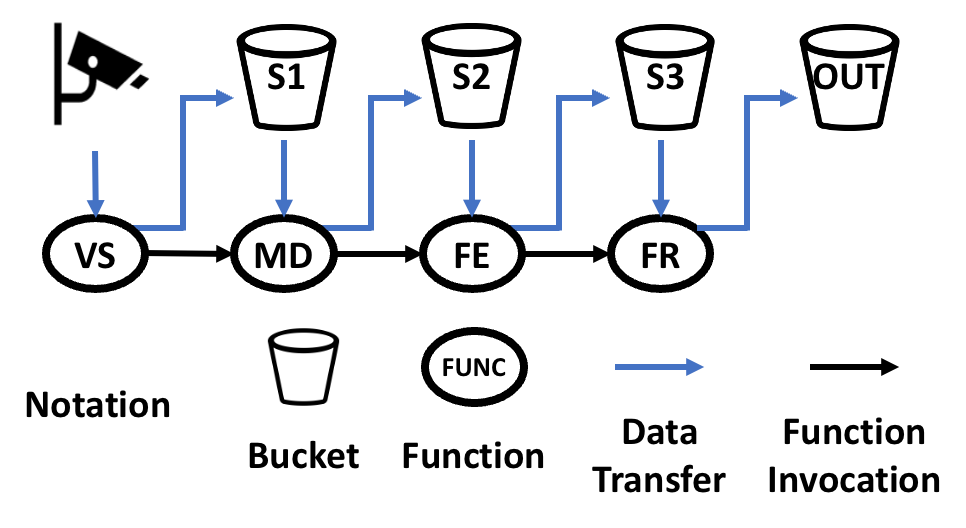}
    \vspace{-12pt}
	\caption{\small Video analytics workflow consisting of four functions: video-splitting (VS), motion-detection (MD), face-extraction (FE), and face-recognition (FR).}
	\label{fig:video-analytics-workflow}
	\vspace{-6pt}
\end{figure}

The video analytics workflow includes functions to process and analyze videos captured by cameras on IoT devices, as illustrated in Fig.~\ref{fig:video-analytics-workflow}. These functions run on the video streams in a pipelined fashion, where each function performs one distinct stage of processing on the previous stage's output. 

The camera records a 2-seconds video. The \textit{video-splitting} function 
divides the 2-seconds video into frames and organizes them into Groups of Pictures (GoP) using FFmpeg. The \textit{motion-detection} function analyzes the frames to detect any motion by comparing consecutive frames using OpenCV, and saves only those frames where it identifies the motion. The \textit{face-extraction} function then processes these motion-containing frames to detect faces within the frames using  Single Shot MultiBox Detector (SSD)~\cite{liu2016ssd} and extracts facial features using dlib. Next, the \textit{face-recognition} function takes the results from the previous function and uses a pre-trained ResNet-34 model~\cite{resnet34} to encode each detected face. It then classifies the faces using k-nearest neighbors (k-NN) and stores the names of each identified face in a text file.

This workflow can be found in many edge applications, such as video surveillance~\cite{myneni2022scvs} and self-driving cars. It also represents a broader range of edge applications that process streaming IoT data using a multi-staged pipeline.
\nocite{waterfall}

Edge computing enables in-situ or in-transit video processing and analysis to reduce latency, lower costs, and provide time-sensitive responses. However, the limited capabilities of edge resources also require careful placement of the functions across IoT, edge, and cloud tiers to meet the resource demands and performance requirements.
EdgeFaaS allows users of such a pipeline workflow to flexibly explore how to deploy the various stages of the pipeline across IoT, edge, and cloud resources, which we will demonstrate in Section~\ref{sec:va-evaluation}.
List~\ref{listing:va-workflow-creation} presents an example of the workflow configuration.

\subsection{Hierarchical Federated Learning Workflow} 
\label{sec:federated_workflow}

\begin{figure}[h]
	\centering
 	\includegraphics[width=0.65\columnwidth]{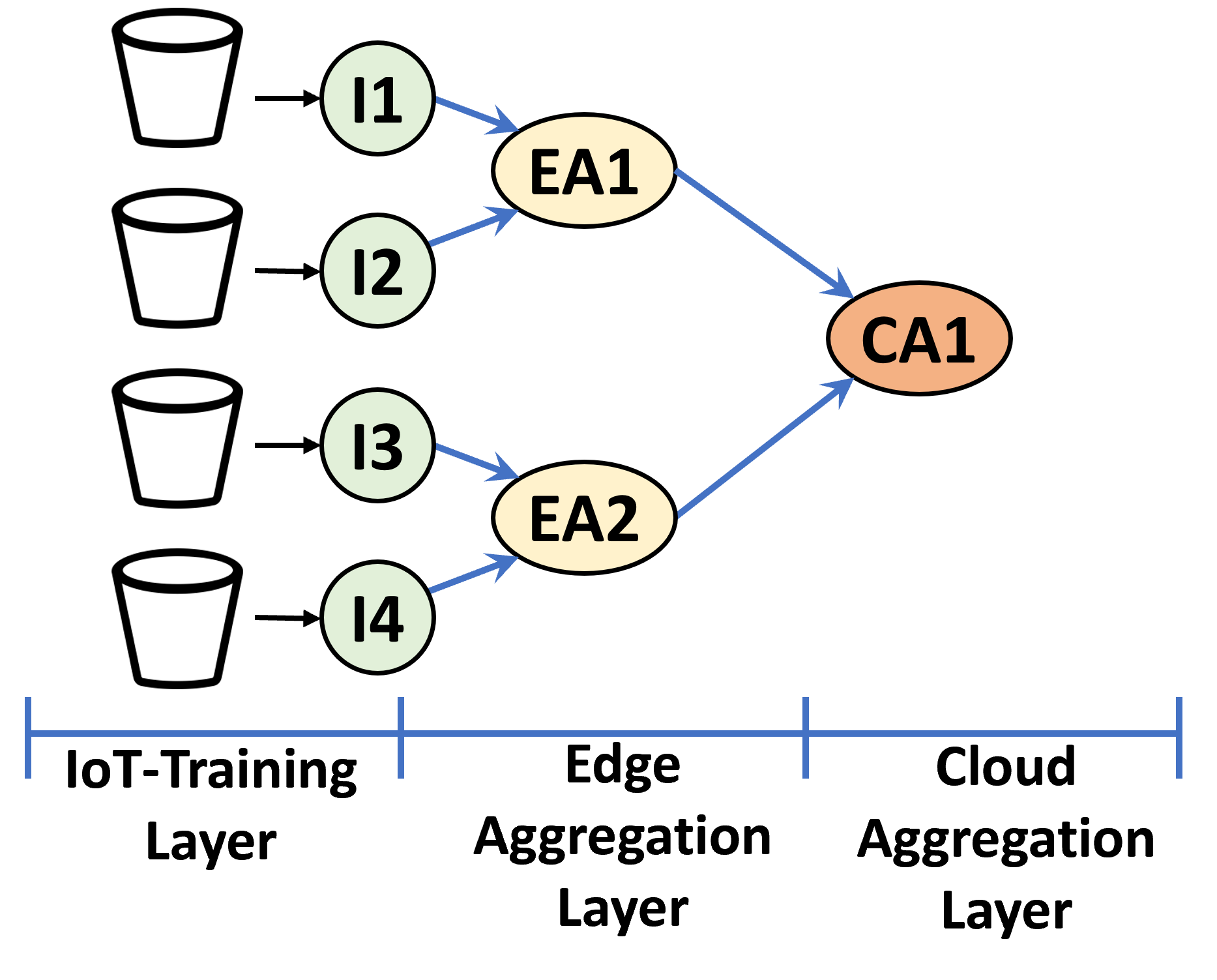}
    \vspace{-12pt}
	\caption{\small Hierarchical federated learning workflow}
	\label{fig:federated-learning-workflow}
	\centering
\end{figure}

The hierarchical federated learning (FL) workflow involves training local data on IoT devices, aggregating the weights of the local models from multiple IoTs on edge resources, and then aggregating the weights of the edge models in the cloud. 
In conventional FL, a central server sends an initial global model to the workers, which trains this model further using its local data, and sends this updated models back to the central server for aggregation to update the global model. This process repeats till the model achieves a satisfactory accuracy. 
In a large system, conventional FL can be slow due to the large number of workers it needs to aggregate and the high communication and synchronization overhead between the workers and the centralized aggregation server.
Hierarchical FL addresses this by following a hierarchical structure for model aggregation, incorporating multiple levels of aggregation servers.  Fig.~\ref{fig:federated-learning-workflow} presents a two-level FL system, which consists of model training workers running on IoT devices, first-level model aggregators running on edge servers, and a second-level aggregator running in the cloud. 
In this way, different clusters of workers can be aggregated asynchronously and each worker needs to interact with only the aggregator in its cluster, thereby improving the scalability and speed of federated learning.

The \textit{iot-training} function's physical deployment is determined by following the data accessibility principle that the necessary input data must be accessible from where the function will be placed. 
In the given example, EdgeFaaS deploys the \textit{iot-training} function on the IoT nodes containing the private buckets (\textit{fl/data1, fl/data2, fl/data3}, and \textit{fl/data4}).
Each instance of the \textit{edge-aggregation} function is deployed on the edge node that is the closest, in terms of average network latency, to \textit{iot-training} functions that it is aggregating. 
Similarly, the \textit{cloud-aggregation} function is deployed at the cloud resource closest to all the \textit{edge-aggregation} functions that it is aggregating.

EdgeFaaS allows users of such a hierarchical aggregation workflow to flexibly adjust the size of worker clusters and explore the tradeoff between learning speed and accuracy as demonstrated in Section~\ref{sec:federated-evaluation}.

\subsection{Audio Classification Workflow}

The audio classification workflow includes functions to classify audio captured by microphones on IoT devices using a machine learning model trained on public data and fine-tuned with local data. 
Audio classification~\cite{audioclassificationsurvey,gupta2025lowpower} is used in many applications, such as wake-word detection (e.g., Amazon Alexa, Apple Siri), environmental sound recognition~\cite{esc-50}, keyword spotting~\cite{kws}, and speaker identification~\cite{speaker-identification}.

This workflow also models the commonly used machine learning workflows that involve model training, model fine-tuning, and model inference.
Typically, a deep learning model is trained on a large, labeled dataset and stored as the generalized pre-trained model. To adapt a model to a new task or new data, fine-tuning adjusts the pre-trained base model using a smaller, task-specific dataset while retaining previously learned features. 
It is often done by freezing most of the pre-trained model and updating only the final classification layers. 
The fine-tuned model is then deployed onto IoT and edge devices for inference. 

The audio classification workflow starts from the \textit{training} function in the cloud which trains the generalized model using an existing dataset (e.g., Audio Set~\cite{AudioSet}). As new IoT data is collected and labeled, the \textit{fine-tuning} function uses the new data to fine tune the generalized model on the edge. The fine-tuned model is then deployed onto the IoTs for audio classification.

EdgeFaaS allows users of such a workflow to flexibly adjust the frequency of fine-tuning and explore the tradeoff between accuracy and efficiency.
More frequent fine-tuning improves the accuracy for classifying new samples that the generalized model is not trained on, but it also increases computational and communication overhead. 
EdgeFaaS allows users to explore and optimize this tradeoff as demonstrated in Section~\ref{sec:audio-classification-evaluation}.

\section{Experimental Evaluation}
\label{sec:experimental_evaluation}

In this section, we evaluate EdgeFaaS using the three representative workflows described in the previous section in a real-world edge computing environment with hundreds of  distributed and heterogeneous resources. 

\subsection{Experimental Setup}
\label{sec:experimental-setup}

\begin{figure}[h]
    \vspace{-6pt}
	\centering
	\includegraphics[width=0.8\columnwidth]{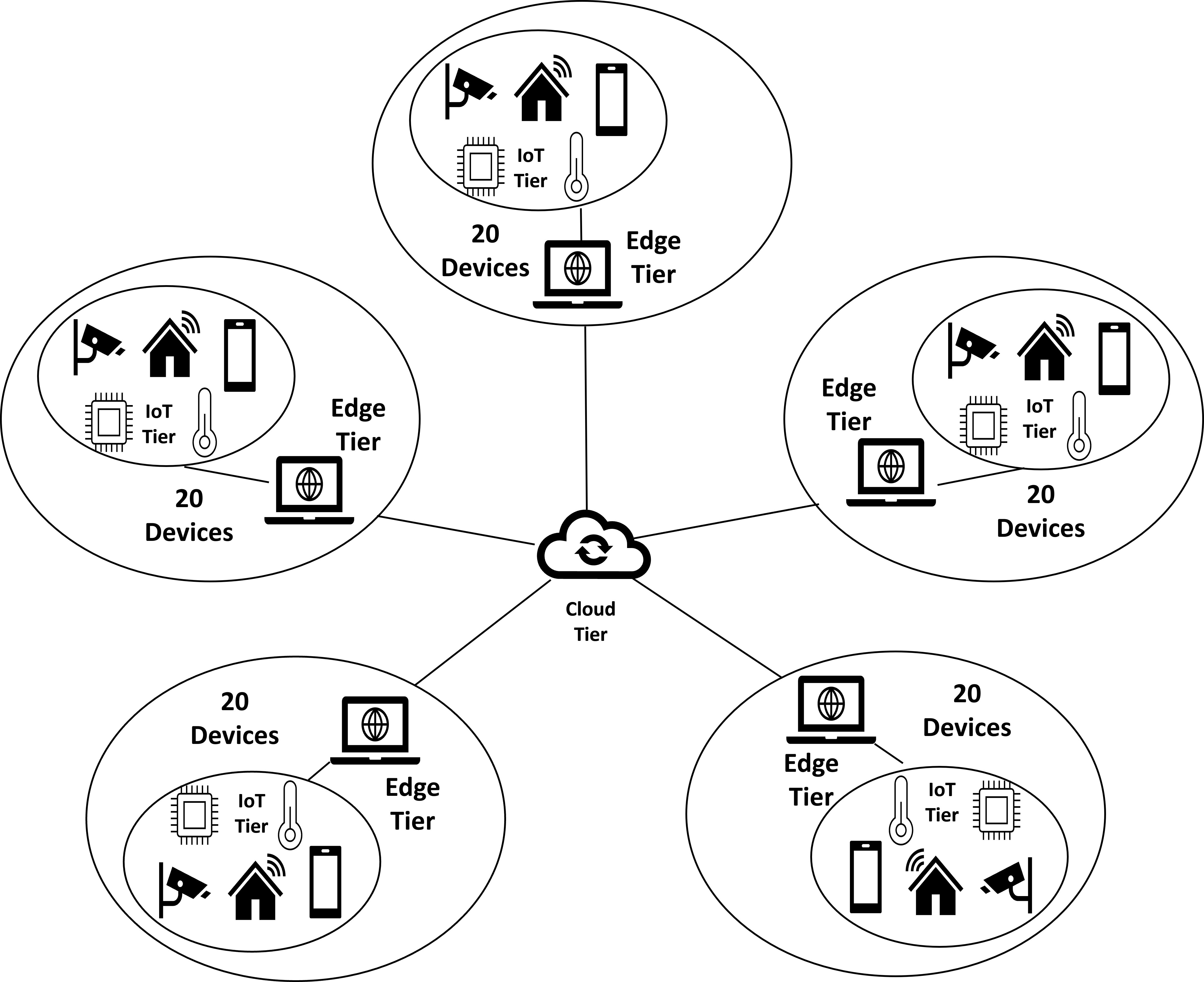}
	\caption{\small \label{fig:setup-geographic} Physical distribution of IoT, edge, and cloud resources in 5 locations for the experimental setup}
\end{figure}

The IoT and edge resources are geographically distributed in five different locations, shown in Fig.~\ref{fig:setup-geographic}. Each location consists of 20 IoT devices and one edge server. For the cloud tier, we use resources from AWS.
Table~\ref{tab:RTT_per_cluster} lists the average round trip time (RTT) between every two tiers in the system.
Table~\ref{tab:eval-hw-spec} lists the details of the resources used in all three tiers including their hardware and software components.

\vspace{-12pt}

\begin{table}[h]
\centering
\caption{Average Round Trip Time (RTT) Comparison Between IoT, Edge, and Cloud Tiers Across Five Locations}
\vspace{-5pt} 
\label{tab:RTT_per_cluster}
\footnotesize
\renewcommand{\arraystretch}{0.8} 
\begin{tabular*}{\columnwidth}{l @{\extracolsep{\fill}} ccccc}
\toprule
\textbf{Locations} & \textbf{1} & \textbf{2} & \textbf{3} & \textbf{4} & \textbf{5} \\
\midrule
IoT-Edge RTT (s)   & 0.02 & 0.02 & 0.09 & 0.03 & 0.04 \\
Edge-Cloud RTT (s) & 0.14 & 0.15 & 0.17 & 0.17 & 0.18 \\
IoT-Cloud RTT (s)  & 0.16 & 0.16 & 0.18 & 0.18 & 0.21 \\
\bottomrule
\end{tabular*}
\end{table}

\vspace{-10pt}

\setlength{\tabcolsep}{3pt}
\begin{table}[h]
\vspace{-12pt}
\centering
\caption{\label{tab:eval-hw-spec} Specification of IoT, Edge, and Cloud resources}
\vspace{-5pt}
\renewcommand{\arraystretch}{0.9} 
\begin{tabular}{p{1.6cm} p{3.2cm} p{2.5cm} p{0.7cm}}
\toprule
\textbf{Resource} & \textbf{Hardware} & \textbf{Software} & \textbf{Tier} \\ 
\midrule
AWS EC2 & \begin{tabular}[c]{@{}l@{}}2.40 GHz Intel Xeon \\ E5-2676 v3,  32 GB \\ Memory, 16 processors\end{tabular} & \begin{tabular}[c]{@{}l@{}}Ubuntu 22.04, \\ OpenFaaS, \\ MinIO, Kubernetes\end{tabular} & Cloud \\ \addlinespace[4pt]
Workstations & \begin{tabular}[c]{@{}l@{}}Intel Core I3, i7, \\ AMD Ryzen 5, 9,\\ 8 - 16 GB Memory \end{tabular} & \begin{tabular}[c]{@{}l@{}}Ubuntu 20.04, \\ OpenFaaS,\\ MinIO, Kubernetes\end{tabular} & Edge \\ \addlinespace[3pt]
Raspberry Pi (3B+, 4, 5) & \begin{tabular}[c]{@{}l@{}}Cortex-A53, A72, A76 \\ Quad-core, 4 GB Memory\end{tabular} & \begin{tabular}[c]{@{}l@{}}Raspberry Pi OS \\ 64-bit Debian 11, \\ Faasd, MinIO\end{tabular} & IoT \\ \addlinespace[3pt]
ESP32-S3 WROOM-2 & \begin{tabular}[c]{@{}l@{}} Xtensa dual-core, 512 KB\\ SRAM, 16 MB PSRAM \end{tabular} & OTA & IoT \\ 
\bottomrule
\end{tabular}
\end{table}

\subsection{Video Analytics Workflow}
\label{sec:va-evaluation}

The workflow consists of four functions: \textit{ video-splitting, motion-detection, face-extraction}, and \textit{face-recognition} which run sequentially.
EdgeFaaS enables users to explore and optimize the tradeoff between computation and communication costs when deploying the various stages of such a pipeline across different tiers.

\subsubsection{Configurations}
\begin{figure}[h]
    \vspace{-6pt}
	\centering
      \includegraphics[width=1\columnwidth]{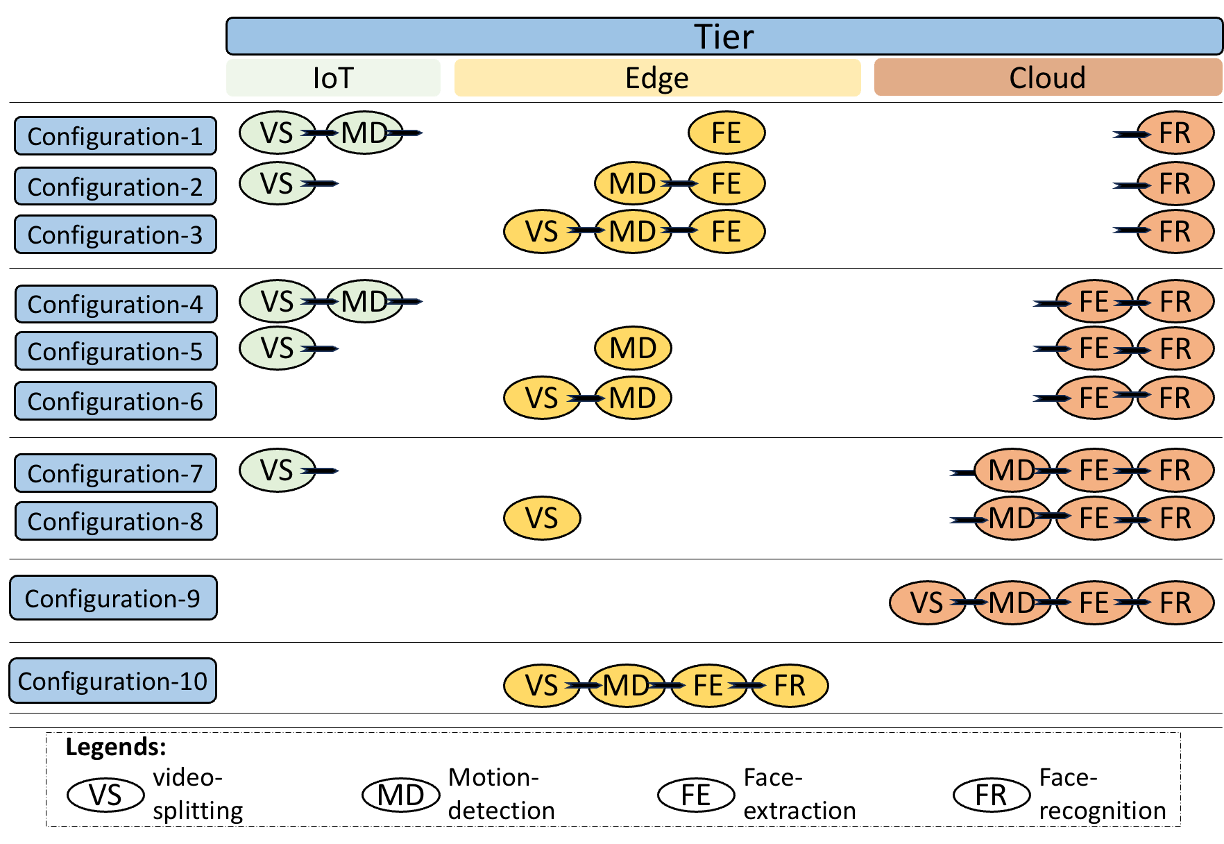}
      \vspace{-18pt}
	\caption{\small \label{fig:VA-configurations-evaluation} Configuration settings for video analytics workflow }
	\centering
    \vspace{-3pt}
\end{figure}

Fig.~\ref{fig:VA-configurations-evaluation} shows all the possible configurations of placing the four functions of the video analytics workflow across IoT, edge, and cloud tiers.
Due to their complexity, \textit{face-extraction} and \textit{face-recognition} functions cannot be deployed on the IoT tier implemented by Raspberry Pi.

\subsubsection{Compute time comparison}

\begin{figure}[h]
    \vspace{-6pt}
	\centering
      \includegraphics[width=0.85\columnwidth]{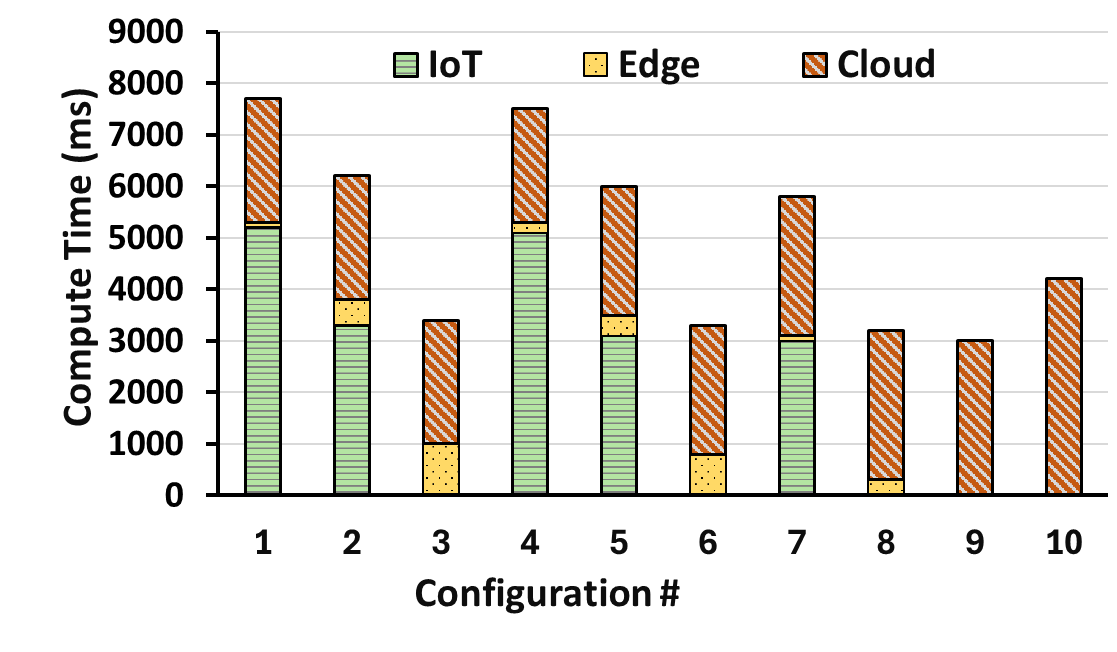}
      \vspace{-12pt}
	\caption{\small \label{fig:VA-compute} Compute time for video analytics workflow}
	\vspace{-9pt}
\end{figure}

Fig.~\ref{fig:VA-compute} compares the total compute time of all the functions across different configurations. When all functions are deployed on the cloud tier, the face-recognition (FR) function consumes the most time at 75\%, followed by motion-detection (MD) at 13\%, video-splitting (VS) at 10\%, and face-extraction (FE) at 2\%. 

As shown in Fig.~\ref{fig:VA-compute}, compute time decreases as more functions run on the cloud tier. Configuration 9, with all functions in the cloud, has the fastest compute time (3.09 sec). 
Compute time significantly increases as more functions are moved to the IoT tier. With two functions on IoT, Configurations 1 and 4 are 60.9\% slower than Configuration 9.

\subsubsection{Communication time comparison}
\begin{figure}[h]
    \vspace{-6pt}
	\centering
      \includegraphics[width=0.85\columnwidth]{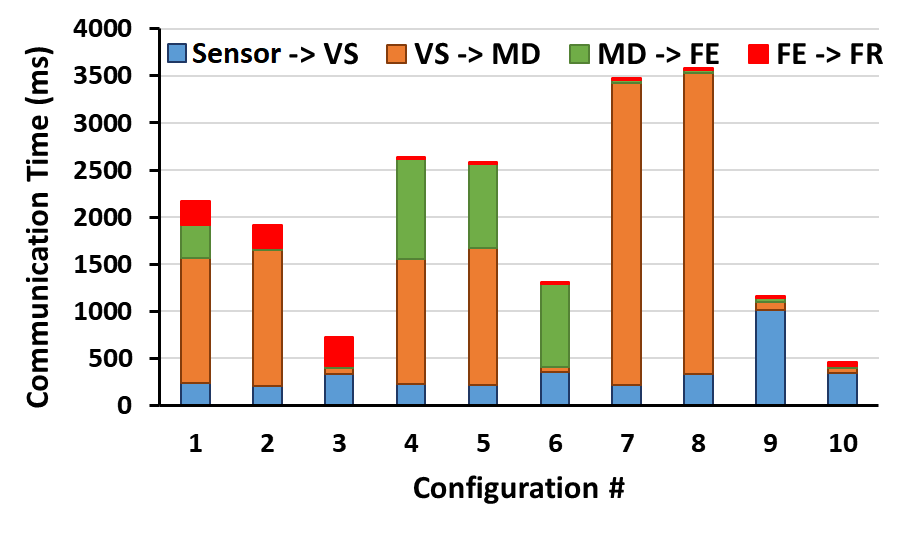}
    \vspace{-12pt}
	\caption{\small \label{fig:VA-communication} Communication time for video analytics workflow }
	\vspace{-6pt}
\end{figure}

Fig.~\ref{fig:VA-communication} compares the communication time, which is mainly the time to transfer data between functions, across the configurations.
The communication time is mainly influenced by data size and network distance. The largest data transfer is from VS to MD (4100 KB), followed by MD to FE (172.7 KB), and finally from FE to FR (6.9 KB).
When all the functions are co-located on the edge tier, as in Configuration 10, the total communication is the lowest. 
Because the VS to MD data transfer is the largest in the entire workflow, we observe that keeping VS and MD on the same tier helps bring down the total communication time. The exception is when VS and MD are co-located on the IoT tier; the weak hardware of the IoT device still makes the local data transfer as expensive as when they are placed on different tiers. 

\subsubsection{Compute vs. Communication time}

\begin{figure}[h]
    \vspace{-6pt}
	\centering
      \includegraphics[width=0.85\columnwidth]{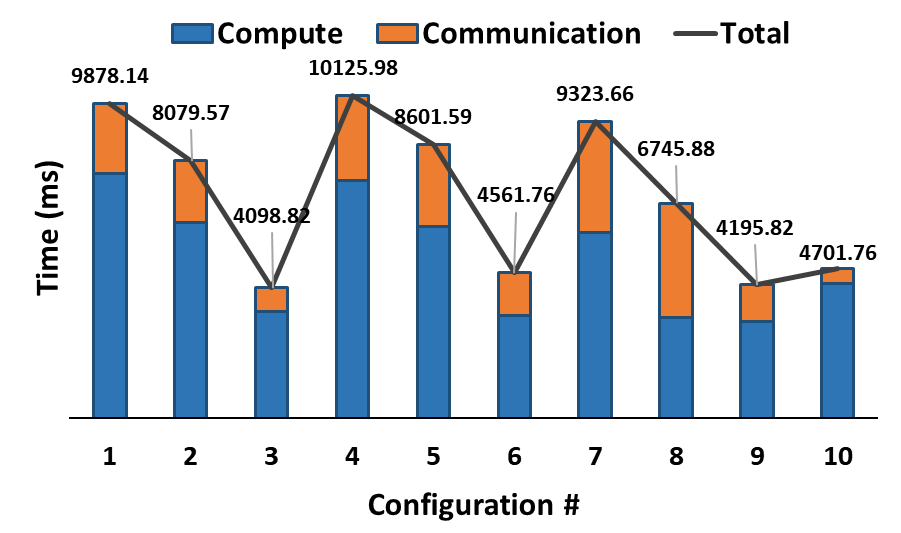}
      \vspace{-12pt}
	\caption{\small \label{fig:VA-compute_communication} End-to-end latency of the video analytics workflow }
	\vspace{-6pt}
\end{figure}

Fig.~\ref{fig:VA-compute_communication} illustrates the end-to-end latency of the video analytics workflow as the summation of the compute time and communication time of all the functions. Although Configuration 9 has the lowest compute time and Configuration 10 achieves the lowest communication latency, Configuration 3 delivers the lowest end-to-end latency by best balancing computation and communication for this workflow. Specifically, co-locating the first three stages on the edge tier eliminates most inter-stage network communication, while offloading the final stage to the cloud substantially reduces computation cost.
This experiment demonstrates EdgeFaaS' ability to allow users to explore the compute vs. communication time tradeoff and find the optimal placement for such a pipeline.

\subsection{Hierarchical Federated Learning Workflow}
\label{sec:federated-evaluation}

The workflow consists of three functions: \textit{ iot-training, edge-aggregation}, and \textit{cloud aggregation} which run on the IoT layer, edge layer, and cloud layer, respectively. EdgeFaaS allows users to explore different resource clustering configurations and explore the tradeoff between the training speed and model accuracy. This experiment involves 100 geographically distributed IoT devices, so it also showcases EdgeFaaS' scalability.

\subsubsection{Model and Dataset}

The workflow trains a LeNet-5 CNN model for handwritten digit recognition. It includes an input layer for 32x32 pixel grayscale images, two 2D convolution layers with ten 5x5 filters each, a dropout layer, and two fully connected layers. The final layer maps to an output layer with ten classes, each representing a digit. It uses the MNIST dataset~\cite{deng2012mnist} containing a training set of 60,000 and a test set of 10,000 examples. We split the dataset among the 100 IoT devices in the experiment, which are implemented by Raspberry Pis. Each IoT device assumes its allocated split is its private data, and uses this data as the input to the \textit{iot-training} function that runs on the device to train its local model. The hyperparameters are training batch size of 64, testing batch size of 1000, learning rate of 0.1, and momentum of 0.5. 
Every IoT device sends its trained local model to the closest edge server as the input to the \textit{edge-aggregation} function. After the edge aggregation is done, the resulting model is sent to the cloud tier as the input to the \textit{cloud-aggregation} function.

\subsubsection{Configurations}

\begin{figure}[h]
    \vspace{-6pt}
	\centering
	\includegraphics[width=0.95\columnwidth]{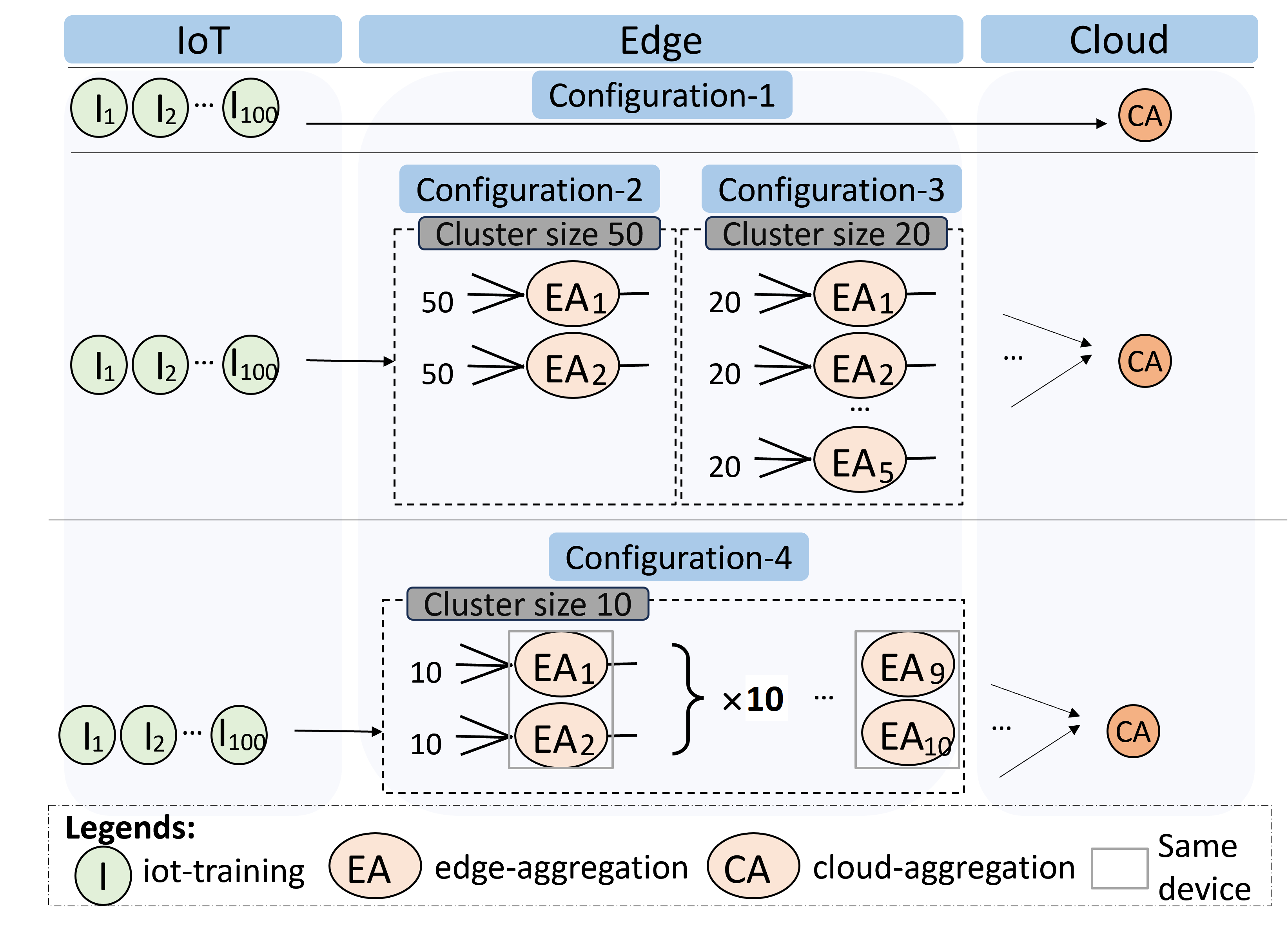}
	\vspace{-6pt}
    \caption{\small \label{fig:FL-configurations-evaluation} Configuration settings for federated learning workflow. Configuration 1 illustrates traditional federated learning involving two tiers in a single-cluster scenario. Configurations 2, 3, and 4 depict hierarchical federated learning across three tiers, with cluster sizes of 50, 20, and 10, respectively.}
	\centering
    \vspace{-3pt}
\end{figure}

As illustrated in Fig.~\ref{fig:FL-configurations-evaluation}, Configuration 1 represents traditional non-hierarchical FL, where all IoT devices form a single cluster and their local models are aggregated by a single aggregator running in the cloud; the other three configurations are hierarchical FL with different cluster sizes and number of clusters. 
Every cluster of IoT devices connects to one edge server selected based on the shortest average distance to the cluster.

To provide a fair comparison of training time, we consider the time that it takes to aggregate the models trained by the IoT devices for the same number of iterations (2,500), which is sufficient for the model to converge. We also let each IoT device train 100 iterations before uploading its model for aggregation; and each edge aggregator (if used) aggregates 5 times before uploading its model for aggregation. Under these settings, with Configuration 1, the cloud aggregator aggregates 25 times; with Configurations 2, 3, and 4, the cloud aggregator aggregates 5 times.

\subsubsection{End-to-end latency comparison}

\begin{figure}[h]
    \vspace{-6pt}
	\centering
	\includegraphics[width=0.8\columnwidth]{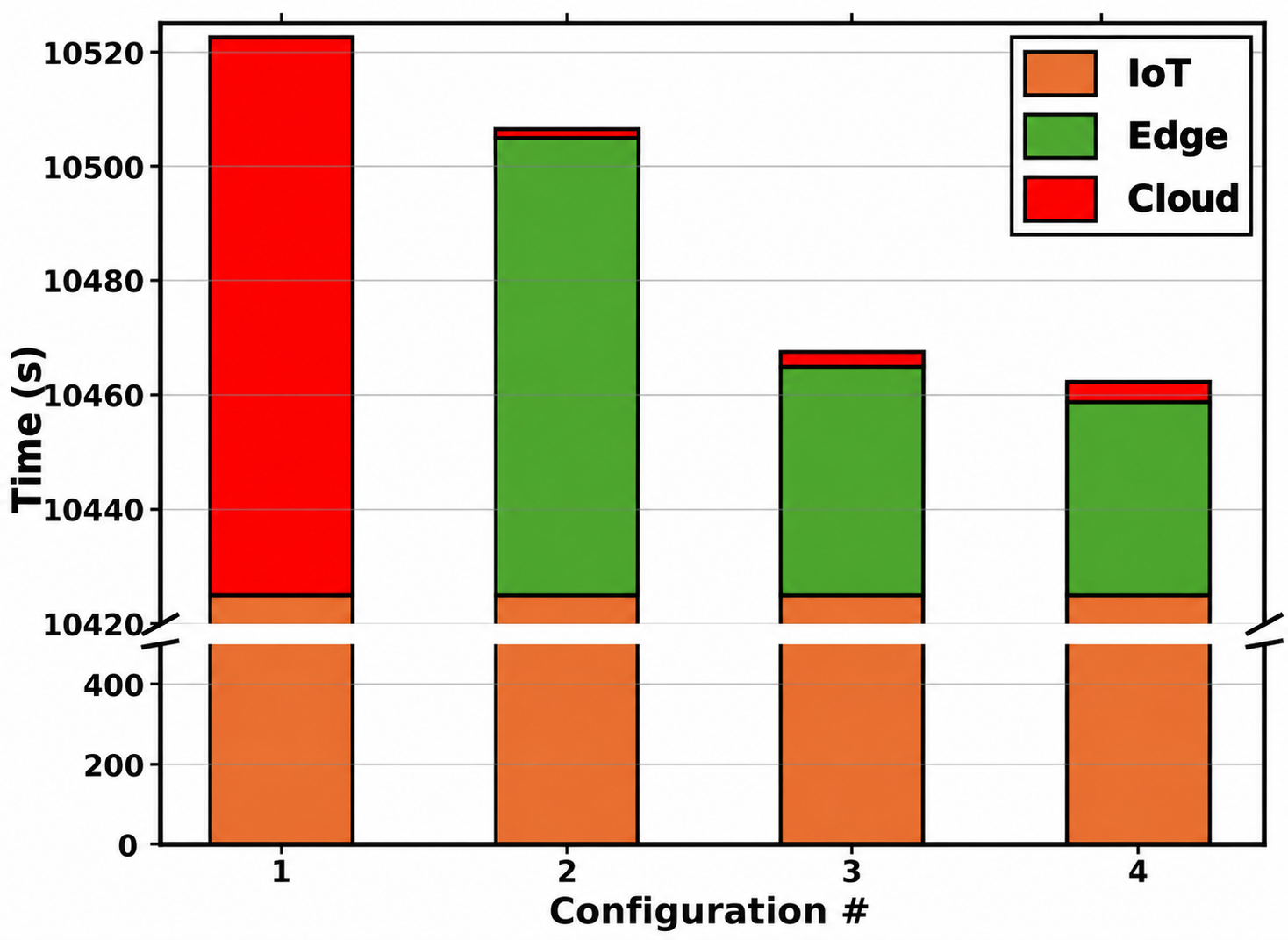}
    \vspace{-9pt}
	\caption{\small \label{fig:total-time-one-round}Total training time of the federated learning workflow for the different configurations}
	\vspace{-3pt}
\end{figure}

We measure the total training time of the different configurations and compare them in Fig.~\ref{fig:total-time-one-round}. For Configuration 1, the measurement includes the time for IoTs to download/upload the model from/to the cloud and train the model locally (\textit{IoT}) and for the cloud to aggregate the models (\textit{Cloud}). For Configurations 2, 3, and 4, the measurement includes the time for IoTs to download/upload the model from/to the edge servers and train the model locally (\textit{IoT}), for the edge servers to download/upload the model from/to the cloud and aggregate the IoT models (\textit{Edge}), and for the cloud to aggregate the edge models (\textit{Cloud}).

The results show that non-hierarchical FL is the slowest to get the global model
because the cloud aggregator has to wait for all of the 100 IoT devices to upload their local models, and all the IoT devices have to download the aggregated model from the cloud. In comparison, hierarchical FL reduces training time by allowing each edge aggregator to work with a smaller number of IoT devices in its proximity, which reduces both synchronization time and model transfer time; and with decreasing cluster size, the training time further reduces.

\subsubsection{Accuracy comparison}

\begin{figure}[h]
    \vspace{-9pt}
	\centering
	\includegraphics[width=0.9\columnwidth]{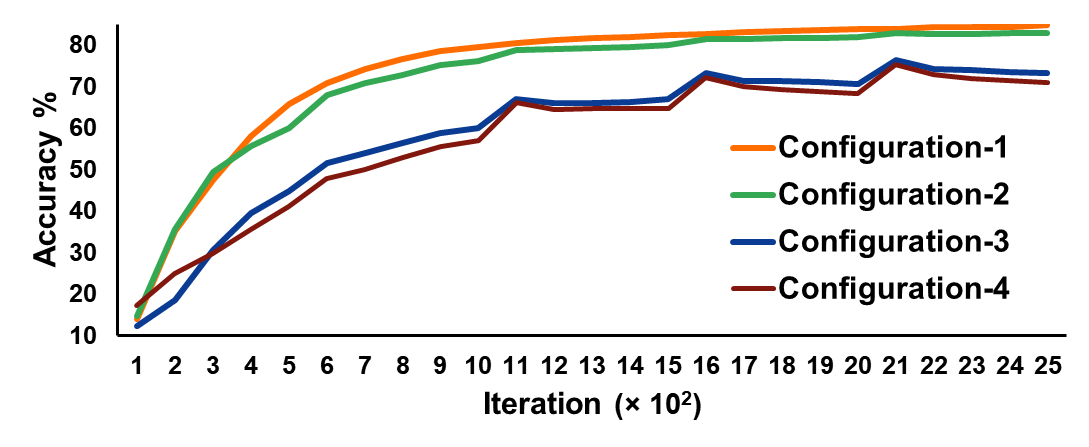}
    \vspace{-12pt}
	\caption{\small \label{fig:FL-accuracy_allConfigs} Accuracy of federated learning for different configurations}
    \vspace{-0pt}
\end{figure}

Fig.~\ref{fig:FL-accuracy_allConfigs} illustrates the model's test accuracy over 2500 training iterations for the four federated learning configurations.
Traditional, non-hierarchical FL shown in Configuration 1 achieves the highest accuracy of 84.74\% over 2500 iterations. Comparing all the hierarchical FL configurations, accuracy reduces with smaller cluster sizes. 
With 10 IoT devices per cluster and a total of 10 clusters, Configuration 4 shows the lowest accuracy at 70.96\%. 
Similar differences can be observed in convergence speed---how fast a model reaches its final accuracy. 
This difference in model accuracy is because, with a larger cluster size, the models in the same cluster are trained with more samples from more IoT devices, which leads to a more generalized global model. 

When Fig.~\ref{fig:total-time-one-round} and Fig~\ref{fig:FL-accuracy_allConfigs} are considered together, this experiment demonstrates that EdgeFaaS allows users to make the tradeoff between training speed and accuracy.

\subsection{Audio Classification Workflow}
\label{sec:audio-classification-evaluation}

The workflow consists of three functions: \textit{training}, \textit{ fine-tuning}, and \textit{inference} which run on the cloud, edge, and IoT resources, respectively. In this experiment, the IoT devices are implemented by ESP32 microcontrollers. EdgeFaaS allows users to explore different fine-tuning frequencies  and analyze the tradeoffs between accuracy and efficiency.

\subsubsection{Model and dataset}

The \textit{training function} trains the YamNet~\cite{YamNet_blog} model on the AudioSet~\cite{AudioSet} dataset. 
YamNet uses a depthwise separable MobileNetV1 convolution architecture that accepts a 1-D float32 Tensor or NumPy array containing a waveform of arbitrary length, represented as single-channel (mono) 16 kHz samples in the range [-1.0, +1.0]. 
The AudioSet dataset is a collection of labeled YouTube excerpts, including 521 distinct audio event categories.  
For fine-tuning, we use 10 classes (ESC-10) from the ESC-50 dataset, which includes 50 categories of environmental sound, to emulate new data collected by IoT devices.
We construct a simple classifier comprising a feature extraction layer for accepting the audio input, which generates mel-spectrogram patches and passes them to two fully connected (dense) layers with ReLU activation functions. During fine-tuning, we freeze the entire YamNet model except for the final classification layers. 
Finally, the fine-tuned model is quantized using int8 and deployed onto the microcontroller through OTA update.

\subsubsection{Configurations}

We emulate a practical deployment scenario where new categories of data are gradually collected and labeled over time. If the audio classification model is immediately fine-tuned with the new data, it will be able to classify audio belonging to the new categories, but fine-tuning incurs computational cost.
To evaluate the tradeoff between accuracy and efficiency, we design three configurations for the experiment. Configuration 1 fine-tunes the model only once at round 10 when all 10 new categories of data are available.
In Configuration 2, the audio classification model is fine-tuned as soon as a new category becomes available.
In Configuration 3, the model is fine-tuned after two new categories become available.
In Configuration 4, the model is fine-tuned after five new categories become available.

\subsubsection{Accuracy comparison}

\begin{figure}[h]
    \vspace{-6pt}
	\centering
    \includegraphics[width=1.0\columnwidth]{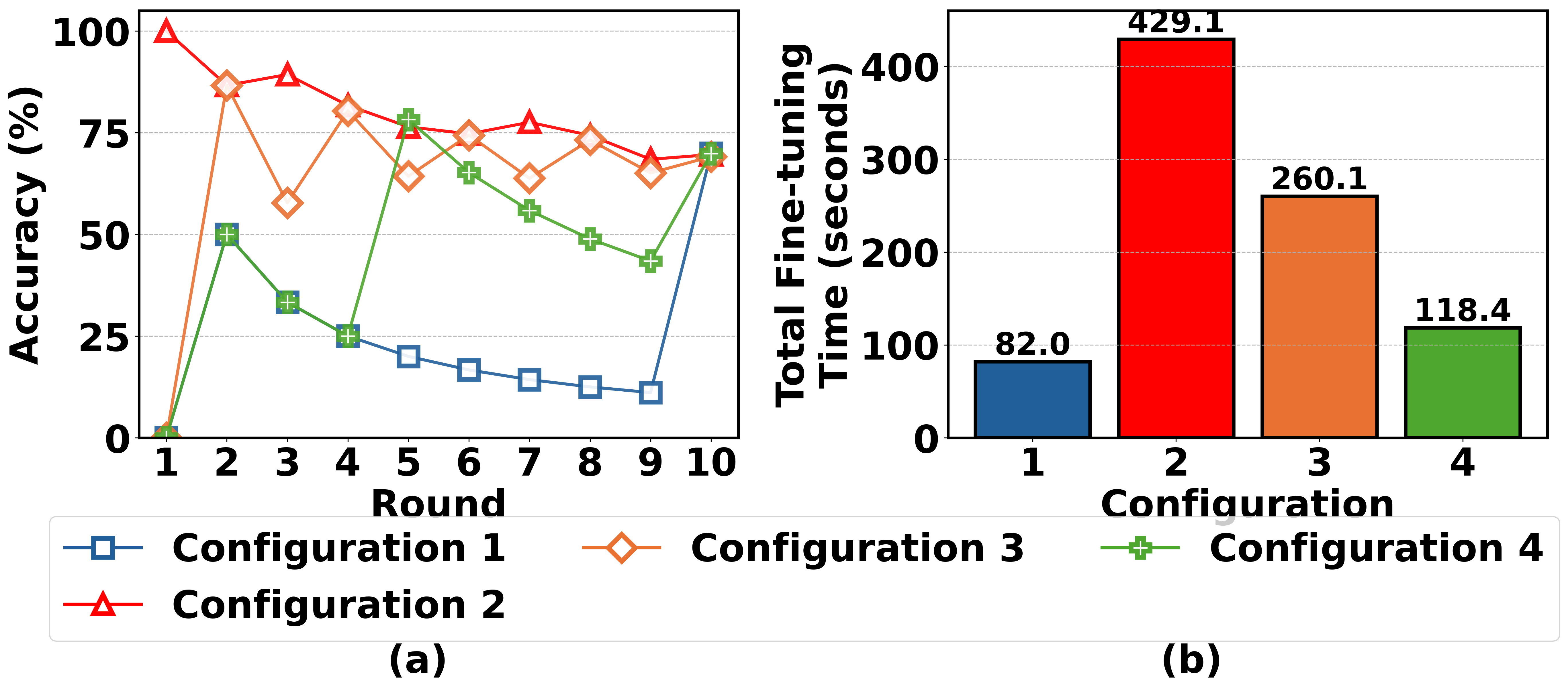}
    \vspace{-18pt}
	\caption{\small (a) Accuracy and (b) fine-tuning time of the audio classification workflow for different configurations.}
    \label{fig:AC_allConfigs}
	\centering
\end{figure}

Fig.~\ref{fig:AC_allConfigs} (a) illustrates the test accuracy trends across different fine-tuning configurations. Configuration 1, which fine-tunes only once at round 10, yields a final accuracy of 69.96\% but performs poorly in earlier rounds with a baseline of just 14.29\%. Configuration 2, which performs fine-tuning in every round by introducing one new class per round, achieves significantly higher accuracy in intermediate rounds compared to Configuration 1. 
Configurations 3 and 4 fine-tune the model less frequently than Configuration 2 and achieve a lower accuracy correspondingly.

\subsubsection{Fine-tuning time comparison}
Fig.~\ref{fig:AC_allConfigs} (b) shows the total fine-tuning time required for the 12 rounds across different configurations. Configuration 1 performs fine-tuning only once at round 10, resulting in the lowest total fine-tuning time of 82 seconds. However, this delayed update deprives users of accurate predictions in earlier rounds, as reflected in the low baseline accuracy. In contrast, Configuration 2 fine-tunes in every round (10 times), leading to the highest total fine-tuning time of 429.1 seconds, which is 5X higher than Configuration 1.
The fine-tuning time of the other configurations are in between these two. 

Taking both the fine-tuning overhead and classification accuracy into consideration, users can decide the best fine-tuning frequency based on their needs. EdgeFaaS enables such flexible exploration by allowing users to conveniently adjust the workflow configuration and evaluate it on distributed, heterogeneous resources.

\subsection{Orchestration overhead and scalability}

We evaluated the overhead and scalability of EdgeFaaS' distributed orchestration under function workloads with different invocation rates. The system employed one global orchestrator and three edge orchestrators. Each request invokes a workflow of three parallel functions. To reveal the worst-case overhead from orchestration, the functions do not perform any computation, and their containers are warm. 

\begin{figure}[h]
	\centering
    \includegraphics[width=1.0\columnwidth]{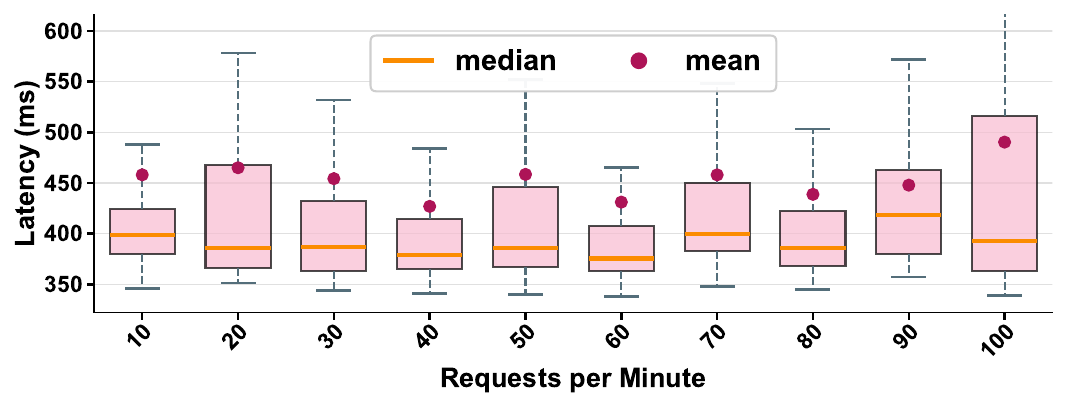}
    \includegraphics[width=1.0\columnwidth]{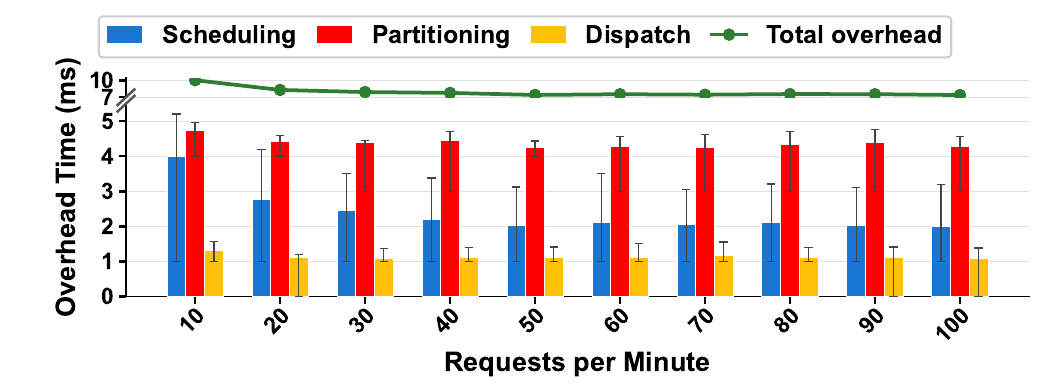}
    \vspace{-18pt}
	\caption{\small \label{fig:e6} (a) End-to-End workflow latency and (b) Orchestration overhead breakdown with different invocation rate}
\end{figure}

Fig~\ref{fig:e6} (a) shows the end-to-end workflow latency, which includes mainly the orchestration overhead. 
The results show that EdgeFaaS provides a stable median latency of 376-418 ms across the entire range of workflow invocation rate, a 10× increase in arrival pressure with no measurable queuing effect. 
This stability is because EdgeFaaS' parallel partition dispatch and per-workflow orchestration introduce no cross-workflow queuing penalty.

Fig~\ref{fig:e6} (b) shows the breakdown of the orchestration overhead comprising scheduling, partitioning, and partition dispatch to edge orchestrators for different invocation rates of the workflow. It drops from 10.5 ms at rate 10 to a stable range of 7.4–7.9 ms beyond rate 30, constituting less than 2.2\% of the total workflow latency. 
The increased overhead at low rates is due to cold-path cache misses in the resource scheduling pipeline, which resolve as the resource and availability caches warm up beyond rate 30. 
Partitioning consistently accounts for the largest share than scheduling time due to additional entry and exit node identification in partitions and cross-partition dependency computation, and incurs an extra write to persist partition descriptors for edge orchestrators to consume. Nevertheless, EdgeFaaS' total overhead remains consistently low as the workflow's invocation rate increases by 10-fold.

\section{Conclusions and Future Work}
\label{sec:conclusion}

EdgeFaaS is a novel function-based framework for edge computing. It virtualizes users, functions, and storage in order to provide unified interfaces for convenient and consistent access to heterogeneous resources distributed across IoT, edge, and cloud tiers.
It also provides isolation for function executions and data accesses, ensuring they operate within their designated namespaces without impacting one another. EdgeFaaS supports an edge application's performance and privacy needs by deploying functions based on their dependencies and storing data only on user-trusted devices.
EdgeFaaS ensures usability by following standards and widely-used protocols for workflow/function specifications and interface designs.
EdgeFaaS is an open-source framework and can enable a variety of follow-up works on many important aspects of edge computing. In particular, the current EdgeFaaS framework employs only basic principles for function and data placement, whereas more advanced function scheduling and data caching and tiering solutions are conceivable. 
EdgeFaaS is a powerful framework for enabling such important studies.

\newpage

\bibliographystyle{IEEEtran}
\bibliography{main_v3}

\end{document}